\documentclass[aps,pre,twocolumn,superscriptaddress,notitlepage]{revtex4-2}
\usepackage{amsmath}
\usepackage{amsfonts}
\usepackage{amssymb}
\usepackage{lmodern,dsfont}
\usepackage{graphicx}
\usepackage[usenames,dvipsnames]{xcolor}
\usepackage{bm}
\usepackage[english=american]{csquotes}
\usepackage{tikz}
\usepackage{scalerel}

\newcommand{\kb}{k_\text{B}}
\newcommand{\av}[1]{\langle #1 \rangle}
\newcommand{\Av}[1]{\left\langle #1 \right\rangle}
\newcommand{\nn}{\nonumber \\}
\newcommand{\n}{\nonumber}

\newcommand{\grad}{\bm{\nabla}}

\renewcommand{\eqref}[1]{Eq.~(\ref{#1})}

\begin{document}

\author{Andreas Dechant}
\affiliation{Department of Physics \#1, Graduate School of Science, Kyoto University, Kyoto 606-8502, Japan}
\author{Eric Lutz}
\affiliation{Institute for Theoretical Physics I, University of Stuttgart, D-70550 Stuttgart, Germany }
\title{Fundamental limits on nonequilibrium sensing}
\date{\today}

\begin{abstract}
The performance of equilibrium sensors is restricted by the laws of equilibrium thermodynamics.
We here investigate the physical limits on nonequilibrium sensing in  bipartite systems with nonreciprocal coupling. We show that one of the subsystems, acting as a Maxwell's demon, can significantly suppress the fluctuations of the other subsystem relative to its response to an external perturbation. Such negative violation of the fluctuation-dissipation relation can considerably  improve the signal-to-noise ratio above its corresponding equilibrium value, allowing the subsystem to operate as an enhanced sensor. We find that the nonequilibrium signal-to-noise ratio of linear systems may be arbitrary large at low frequencies, even at a fixed overall amount of dissipation.
\end{abstract}

\maketitle

Sensing plays a pivotal role in science and technology. By monitoring changes in the surroundings and reacting to external signals, sensors provide essential information about the environment of a system \cite{fra16,her22,bar23}. However, unwanted stochastic fluctuations fundamentally limit the amount of information that can be acquired. A sensor should  provide a strong response to a signal and, at the same time, be minimally affected by the detrimental influence of noise. An important figure of merit that quantifies this property is the signal-to-noise ratio  which describes how good a  detected signal can be distinguished from the noise \cite{fra16,her22,bar23}. Recent studies  of biochemical networks, in particular of the sensing of chemical  concentrations by biological cells,  have revealed that the breaking of detailed balance away from equilibrium can  enhance sensing performance \cite{gov12,meh12,lan12,tu08,sko13,lan14,gov14,nga20}. These findings suggest that operating sensors far from equilibrium can be of  significant advantage. Yet, the fundamental physical limits on nonequilibrium sensing are still unknown  \cite{aqu16,wol16}.

We here address this crucial issue in the context of Maxwell's demon \cite{mar09,lut15}, using the tools of information thermodynamics \cite{sei12,par15}. By measuring  a system and applying feedback, Maxwell's demon is able to extract work from an equilibrium heat reservoir by breaking the fluctuation-dissipation relation that connects the response to an external field to the equilibrium correlation function of spontaneous fluctuations \cite{kub66,mar08}. We  show that the demon may also enhance the sensing ability of the system by strongly suppressing the random fluctuations of the system at the expense of its own fluctuations. As a consequence, the nonequillibrium signal-to-noise ratio  may not only be improved compared to the equilibrium situation, it can be {arbitrarily} large at low frequencies in linear systems, even at a fixed overall amount of dissipation. This result implies that there is actually no fundamental limit on out-of-equilibrium sensing.

We concretely consider a generic composite system whose state space can be divided into two distinct subsystems that interact with each other. This setup acts as an autonomous Maxwell demon where one subsystem generates information and the other one reacts to it  \cite{bar13,hor14,har14,shi15,hor15,fre21}. It additionally provides a general model for molecular sensors and two-component molecular machines that operate without external measurement and feedback \cite{ehr23}. When the composite system is in a nonequilibrium steady state, created for instance by nonconservative forces,  entropy is dissipated and  detailed balance is broken. In the following, we combine a newly derived local form of the Harada-Sasa relation, that relates the dissipated heat to violation of the fluctuation-dissipation relation \cite{har05,ter05,har06,toy07}, and the second law of information thermodynamics, that extends the entropy balance to include the contribution of the information flow between the  subsystems \cite{hor14,har14,shi15,hor15,fre21}. We show that fluctuations of one subsystem (sensor) can be arbitrary reduced compared to its response, when its dissipated heat becomes negative for a sufficiently large information flow to the other subsystem (demon). Such apparent violation of the second law is at the origin of  enhanced nonequilibrium sensing. We illustrate this generic result with the example of two  overdamped harmonic oscillators with nonreciprocal coupling (Fig.~1).

\textit{Harada-Sasa relation for subsystems.} We begin by deriving a Harada-Sasa relation for coupled subsystems. We consider a composite system consisting of $d$ overdamped degrees of freedom $\bm{z}(t) \in \mathbb{R}^d$ in contact with a viscous equilibrium environment characterized by a temperature $T$ and a friction coefficient $\gamma$, whose dynamics obeys the Langevin equation (we  set $\kb = 1$) \cite{ris89}
\begin{align}
\gamma \dot{\bm{z}}(t) = \bm{f}(\bm{z}(t)) + \sqrt{2 \gamma T} \bm{\xi}(t) \label{langevin} , 
\end{align}
where  $\bm{f}(\bm{z})$ are arbitrary forces acting on the system and $\bm{\xi}(t)$ is a vector of mutually independent Gaussian white noises.
When the forces  are nonconservative (for example, external driving forces or nonreciprocal interactions), the nonequilibrium steady state of the system is  characterized by a positive rate of heat dissipation \cite{sei12}
\begin{align}
\label{2}
\!\!\dot{Q}_\text{diss} \!=\! T \sigma \!= \!\Av{\bm{f}^\text{T} \circ \dot{\bm{z}}} \!= \!\frac{1}{\gamma} \Av{ \Vert \bm{f} - T \grad_z \ln p_\text{st} \Vert^2}_\text{st} \!\geq 0,\!
\end{align}
where $\circ$ is the Stratonovich product and $\av{\ldots}_\text{st}$ denotes the average with respect to the steady-state probability density $p_\text{st}(\bm{x})$.
The quantity $\sigma$ is  the total entropy production rate that represents the increase in entropy of both the system and the environment due to the non-equilibrium nature of the dynamics \cite{sei12}.
We further divide the degrees of freedom into two subsets, $\bm{z} = (\bm{x},\bm{y})$, and interpret $\bm{x}$ and $\bm{y}$ as the degrees of freedom of the subsystems $X$ and $Y$, respectively. Subsystem $X$ will be the sensor whereas subsystem $Y$ will act as the demon.
Doing the same for the forces, $\bm{f} = (\bm{f}^X,\bm{f}^Y)$, we can split the total dissipation into local contributions from $X$ and $Y$,
$
\dot{Q}_\text{diss} = \Av{\bm{f}^{X,\text{T}} \circ \dot{\bm{x}}} + \Av{\bm{f}^{Y,\text{T}} \circ \dot{\bm{y}}} = \dot{Q}_\text{diss}^X + \dot{Q}_\text{diss}^Y 
$.

\begin{figure}
\includegraphics[width=.49\textwidth]{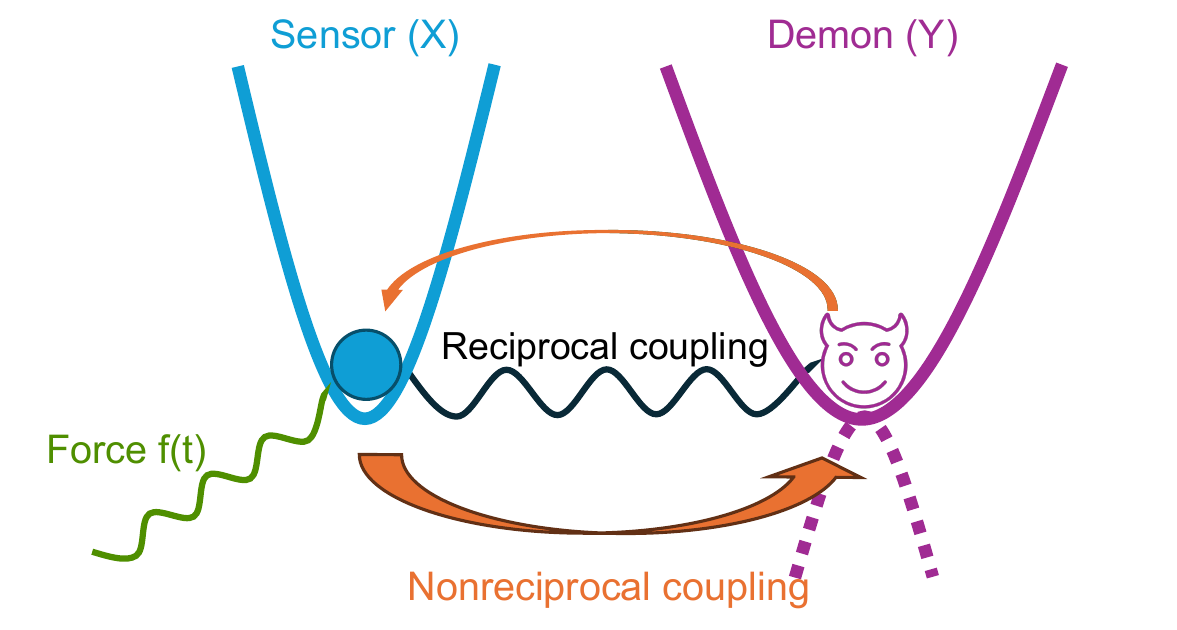}
\caption{Sensor-demon system.
The sensor  consists of a Brownian particle $(X)$  (blue) that is coupled via reciprocal (black spring) and nonreciprocal (orange arrows) interactions to another Brownian particle ($Y$)  (purple) which acts as the demon. The demon increases the nonequilibrium signal-to-noise ratio by reducing the fluctuations of the sensor at the expense of its own. Optimal sensing of a force $f(t)$ (green) is achieved for an inverted harmonic potential for the demon (dotted).} \label{fig-demon}
\end{figure}

Let us next relate the local heat dissipation rates to the local fluctuations and responses of each subsystem.
The fluctuations of the variable $\bm{z}(t)$ can be quantified with  the (positive definite) power spectral density  matrix  \cite{ris89}
\begin{align}
\!\!\!\big(\bm{S}(\omega)\big)_{kl} \!= \!\frac{1}{2}\! \int_{-\infty}^\infty \!\!\!dt \ e^{i \omega t} \!\Av{\delta z_k(t) \delta z_l(0)} \!+ \!\Av{\delta z_l(t) \delta z_k(0)}\!,\!\!
\end{align}
with $\delta \bm{z}(t) = \bm{z}(t) - \Av{\bm{z}}_\text{st}$.
Its integral over all frequencies is equal to the steady-state fluctuations of $\bm{z}(t)$,
$
 \int_0^\infty d\omega \ \big(\bm{S}(\omega) \big)_{kl} /\pi= \Av{\delta z_k \delta z_l}_\text{st} .
$
A closely related quantity is the velocity power spectral density matrix, $\bm{S}_v(\omega) = \omega^2 \bm{S}(\omega)$, which likewise measures the fluctuations of $\dot{\bm{z}}(t)$ in a given frequency interval \cite{ris89}.
On the other hand, the response of $\bm{z}(t)$ to a perturbation force $\epsilon \phi(t) \bm{\hat{e}}_l$ applied in direction $l$ can, to linear order in the magnitude $\epsilon$ of the perturbation, be expressed as \cite{ris89}
\begin{align}
\!\!\!\Av{z_k(t)}^\epsilon - \Av{z_k}_\text{st} \simeq \epsilon \int_0^t \!\!dt' \int_0^{t'}\!\! dt'' \ \mathcal{R}_{v,kl}(t'-t'') \phi(t''),
\end{align}
where $\Av{\ldots}^\epsilon$ denotes the average evaluated in the perturbed system and the matrix $\bm{\mathcal{R}}_v(t'-t'')$ is the velocity-response matrix, whose components measure how much the velocity in direction $k$ at time $t'$ changes in response to an applied force in direction $l$ at time $t''$.
Note that, due to causality, $\bm{\mathcal{R}}_v(t'-t'')$ is only defined for $t' \geq t''$.

To simplify the notation, we proceed by focusing on  a two-dimensional space, $\bm{z} = (x,y)$, with single-variable subsystems.
Then, we can write the two matrices 
\begin{align}
\!\!\!\bm{S}_v(\omega)\! = \!\begin{pmatrix} S_v^X &  S_v^{XY} \\ S_v^{XY} & S_v^Y \end{pmatrix} \text{ and }\bm{R}_v(\omega) \!=\! \begin{pmatrix} R_v^X &  R_v^{XY} \\ R_v^{YX} & R_v^Y \end{pmatrix},\!\!\!
\end{align}
where $S_v^X(\omega)$ and $S_v^Y(\omega)$ are the respective velocity power spectral densities  of $X$ and $Y$, and $S_v^{XY}(\omega) = S_v^{YX}(\omega)$ quantifies the correlations between the two subsystems.
Similarly,  $R_v^X(\omega)$ measures the response of subsystem $X$ to perturbations applied to itself, while $R_v^{XY}(\omega)$ measures the response of subsystem $X$ to perturbations applied to $Y$.
Out of equilibrium, the response is generally not reciprocal, $R_v^{XY}(\omega)\neq R_v^{YX}(\omega)$.

Using the explicit expressions for $\bm{S}_v$ and $\bm{R}_v$, we obtain the local Harada-Sasa relation  for subsystem $X$ (sensor),
\begin{align}
\label{6}
\!\!\frac{\gamma}{\pi} \int_0^\infty \!d\omega \ \big[S^X_v(\omega) - 2 T R^X_v(\omega) \big] = \Av{f^{X} \circ \dot{X}} = \dot{Q}^X_\text{diss}, \!\!
\end{align}
that connects the violation of the local fluctuation-dissipation theorem, $S^X_v(\omega) = 2 T R^X_v(\omega)$ \cite{kub66,mar08}, to the local heat dissipation rate $\dot{Q}^X_\text{diss}$ (Appendix A.4). A similar relation holds for subsystem $Y$ (demon).

\begin{figure*}[t]
    \begin{tikzpicture}
      \node (a) [label={[label distance=-0.2 cm]150: \textbf{a)}}] at (-1,-0.15) {\includegraphics[width=8.2cm]{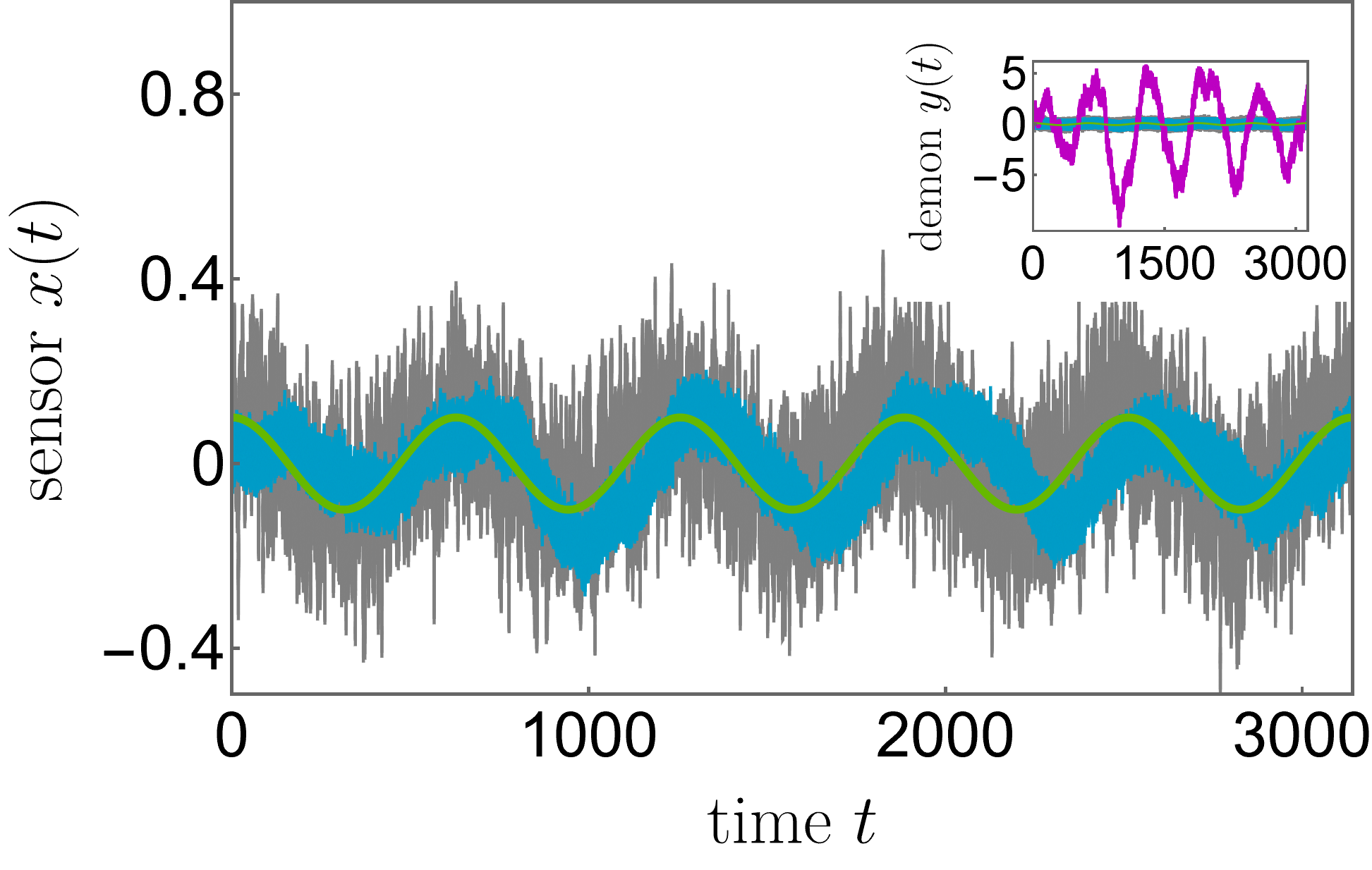}}; 
      \node (b) [label={[label distance=-0.2 cm]150: \textbf{b)}}] at (8,0) {\includegraphics[width=7.7cm]{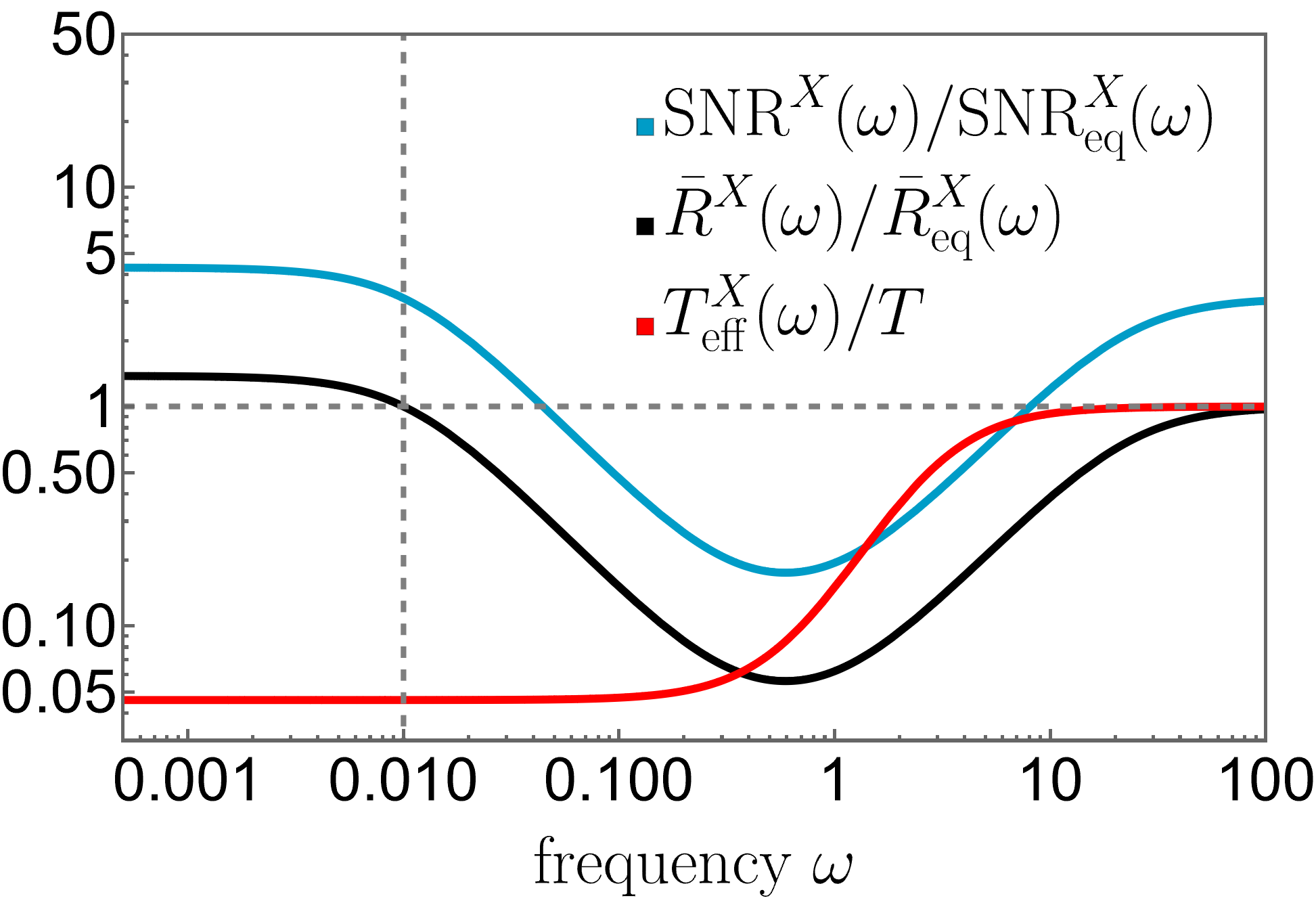}};	
    \end{tikzpicture}
    \caption{Enhanced nonequilibrium sensing. a) The response of the sensor $x(t)$ to a periodic perturbation $\epsilon \cos(\omega_0 t)$ (green) exhibits  less fluctuations in the presence of the demon (blue) than in equilibrium (gray). The inset shows the much larger fluctuations of the demon (purple). b)  Response function $\bar{R}_v^X(\omega)$, \eqref{response-absolute}  (black), and signal-to-noise ratio, \eqref{SNR-definition} (blue),   normalized by their equilibrium values, which they both exceed  below $\omega_0$ (vertical dotted line).  The effective temperature $T_\text{eff}^X(\omega)$ (red) is always below the environment temperature $T$.
 Parameters are $\sigma = 10$, $T = 0.01$, $\omega_0 = 0.01$, $\epsilon=0.1$ and $\gamma = 1$. Coupling  parameters after minimization of the signal-to-noise ratio  for a constant response are $k_x = 15.50$, $k_y = -7.919$, $\kappa = 8.269$, $\delta = -7.766$, corresponding to an eigenvalue $\lambda^- = 0.0105$ of the force matrix.   }
 \label{fig-trajectory}
 \end{figure*}
  
\textit{Improved nonequilibrium sensing.}
Equation (\ref{6}) for the local subsystem has the same form as the global Harada-Sasa relation for the  composite system \cite{har05,ter05,har06,toy07}. However, the underlying physics is radically different. According to the global second law, $  \dot{Q}_\text{diss} \geq 0$, \eqref{2}, the rate of heat dissipation is positive. The Harada-Sasa relation then implies that driving the system out of equilibrium  always reduces the overall response compared to the fluctuations. This seems to suggest that better sensing, with a larger signal-to-noise ratio, is to be achieved near equilibrium. By contrast, the local second law for subsystem $X$ reads $\dot{Q}^X_\text{diss} + T l^X \geq 0 $,
where $ l^X =  \Av{(\bm{f}^X - T \grad_x \ln p_\text{st})^\text{T} \grad_x \ln p_\text{st}}_\text{st}/\gamma$ is the so-called learning rate, which quantifies the information flow between the subsystems \cite{hor14,har14,shi15,hor15,fre21}. Through the action of the demon,
 the local heat dissipation rate $\dot{Q}^X_\text{diss}$ of  subsystem $X$ can become negative in the presence of a sufficiently large information flow $l^X$. This effect allows one to cool $X$ or to continuously extract work from it;  it is the foundation for what has been termed nonreciprocal cooling \cite{xu19,loo20,loo23}. A direct consequence of  \eqref{6} is that the demon, with the help of the same effect, can also suppress the fluctuations of the subsystem   compared to its response, and hence increase the signal-to-noise ratio.
 
The reduced fluctuations may be quantified  with the effective temperature, 
$
T_\text{eff}^X(\omega) = {S_v^X(\omega)}/{2 R_v^X(\omega)}
$,
which measures the magnitude of the fluctuations of $X$ compared to its response \cite{cug11}.
In equilibrium, the effective temperature is equal to the environmental temperature,  $T_\text{eff}^X = T$.
Conversely, a value of $T_\text{eff}^X$ smaller than $T$ indicates that the response  is enhanced compared to the fluctuations, and thus the system can be expected to perform better as a sensor.
However, $T_\text{eff}^X$ may not always characterize the practical usefulness of the system as a sensor.
The reason is that the response function $R_v^X(\omega)$  in \eqref{6} is the real part of the complex response function, and therefore only accounts for the in-phase response of the velocity.
In practice, we are often interested in the amplitude of the response, which is characterized by the absolute value of the complex response function \cite{ris89},
\begin{align}
\bar{R}_v^X(\omega) &= \sqrt{R_v^X(\omega)^2 + \tilde{R}_v^X(\omega)^2} \label{response-absolute}.
\end{align}
The imaginary part, $\tilde{R}_v^X(\omega) =   \int_0^\infty dt \sin(\omega t) \mathcal{R}_v^X(t) 
$, of the complex response function measures the out-of-phase response of the velocity.
Equation (\ref{response-absolute}) can be used to define the dimensionless signal-to-noise ratio  of the sensor
\begin{align}
\text{SNR}^X(\omega) = \frac{\bar{R}_x^X(\omega) f}{ \sqrt{\text{Var}_\text{st}(x)}}= \frac{\bar{R}_v^X(\omega) f}{\omega \sqrt{\text{Var}_\text{st}(x)}} \label{SNR-definition},
\end{align}
where $f$ is the applied perturbation and $\text{Var}_\text{st}(x)$ measures the overall fluctuations of $x$.
The main result of this paper is that there is no fundamental upper limit on $\text{SNR}^X$: in principle, we may design a system that has arbitrarily small fluctuations compared to the response, as we will now demonstrate in a concrete system.

\begin{figure*}
\begin{tikzpicture}
      \node (a) [label={[label distance=-0.2 cm]150: \textbf{a)}}] at (-1,0) {\includegraphics[width=7.7cm]{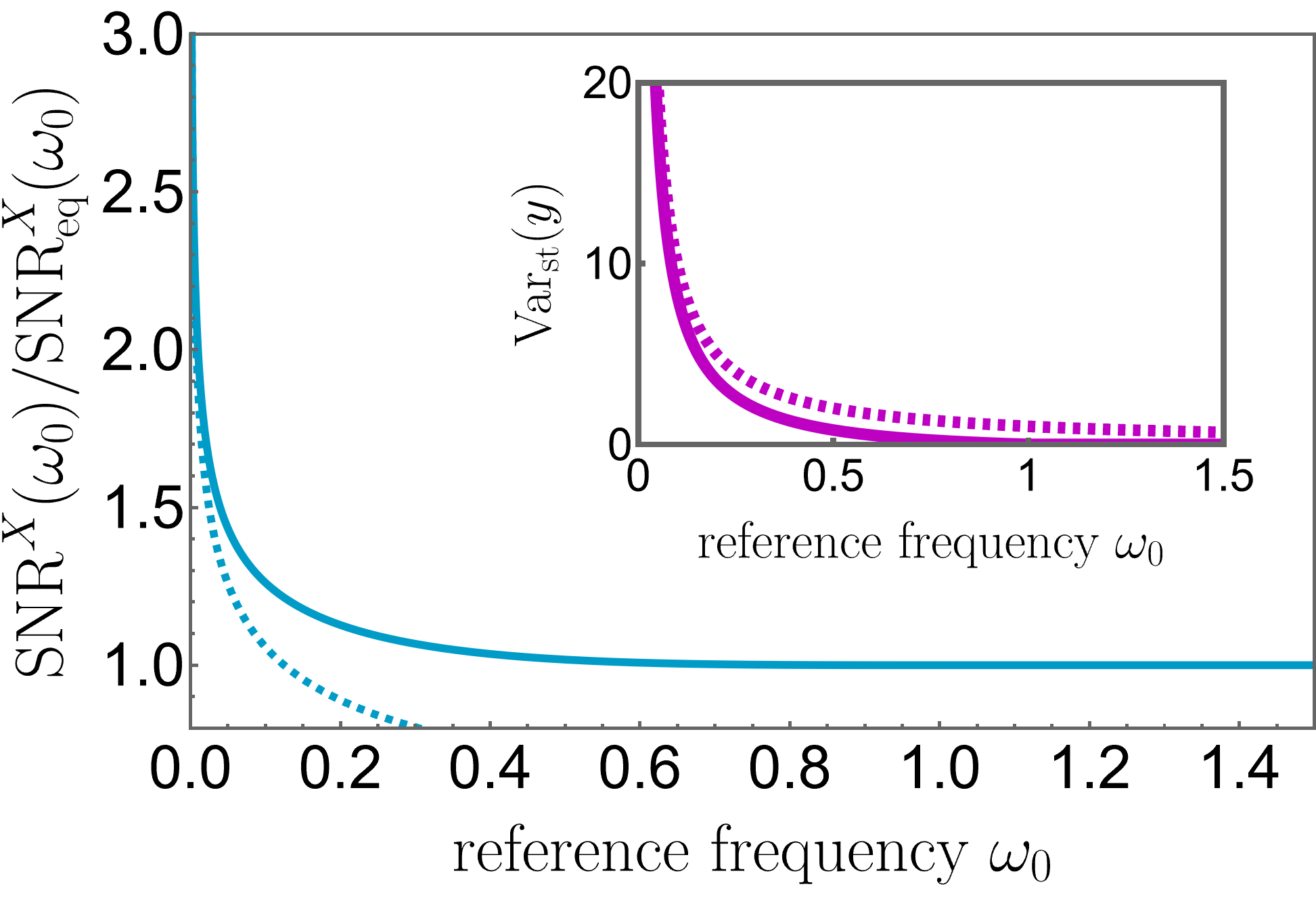}}; 
      \node (b) [label={[label distance=-0.2 cm]150: \textbf{b)}}] at (8,0) {\includegraphics[width=7.7cm]{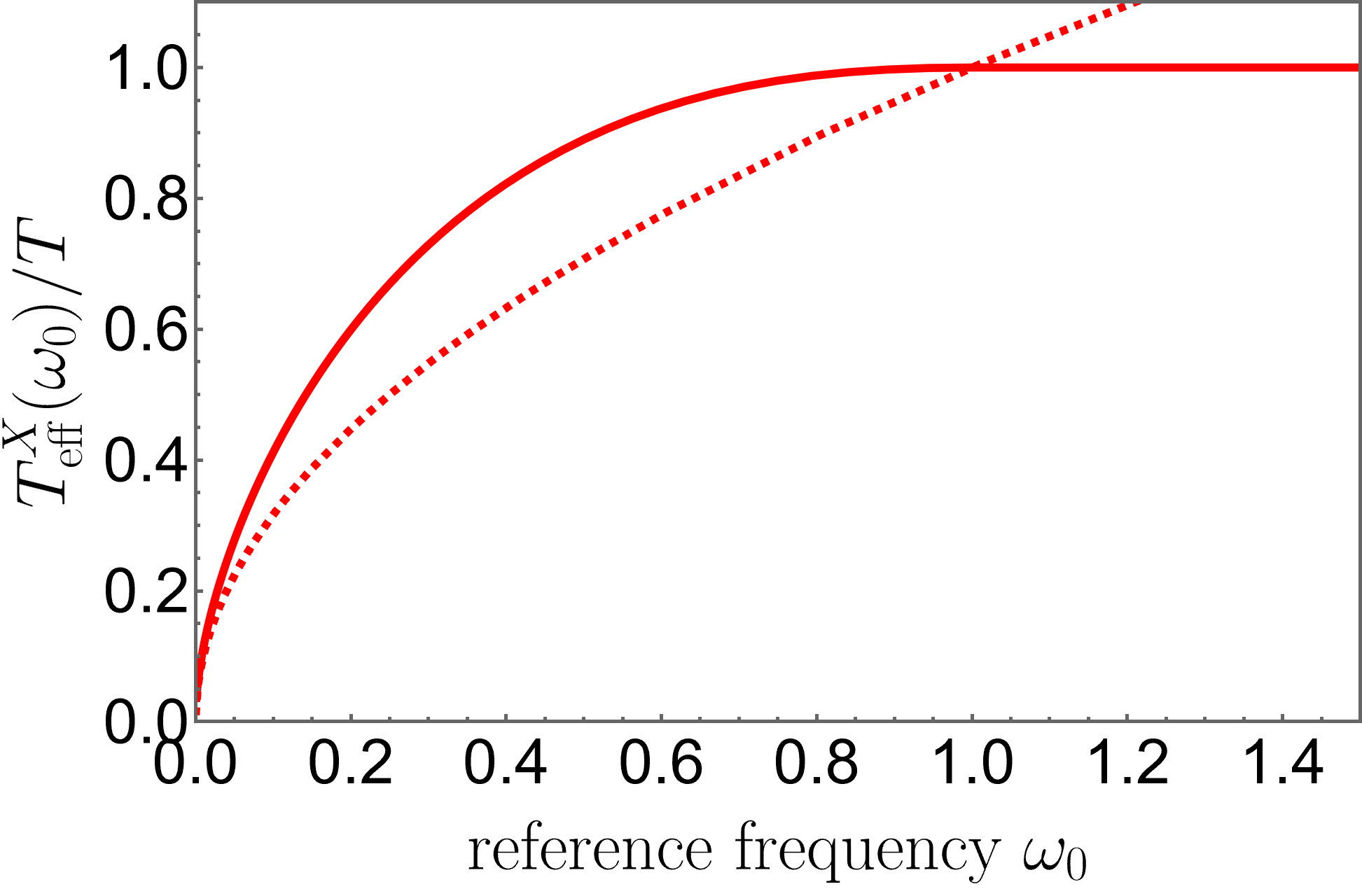}};	
    \end{tikzpicture}
\caption{Nonequilibrium sensing limit. a)  The nonequilibrium  signal-to-noise ratio, \eqref{SNR-definition} (blue solid), exceeds its  equilibrium value  for low frequencies $\omega_0$, and diverges as $\omega_0^{\scaleto{-1/4}{5pt}} $ for $\omega_0\rightarrow 0$, \eqref{asymptotic} (blue dashed), indicating the absence of a fundamental limit on out-of-equilibrium sensing. The result is obtained by fixing  the response to the equilibrium response, $\bar{R}_v^X(\omega_0) = \bar{R}_{v,\text{eq}}^X(\omega_0)$, corresponding to $k_x = 1$, with  $\sigma = 1$, and then minimizing the variance with respect to the eigenvalues $\lambda^+$ and $\lambda^-$ of the force matrix. The divergence of the signal-to-noise ratio of the sensor is accompanied by diverging fluctuations of the demon (purple, inset). b) The  effective temperature  (red solid) vanishes as $\omega_0^{\scaleto{1/2}{5pt}}$ for $\omega_0\rightarrow 0$, \eqref{asymptotic} (red dashed).} \label{fig-optimization}
\end{figure*}

\textit{Application to  a linear system.}
Let us consider  a two-dimensional system where  subsystems $X$ and $Y$ can be locally  approximated by linearly coupled harmonic oscillators (Fig.~1). The dynamics of the composite system follows the Langevin equation (\ref{langevin}) with $\bm{f}(\bm{z}(t)) = - \bm{K} \bm{z}(t)$. We parameterize the force matrix $\bm{K}$ as
\begin{align}
\bm{K} = \begin{pmatrix} k_x + \kappa & -\kappa - \delta \\ -\kappa + \delta & k_y + \kappa \end{pmatrix} .
\end{align}
This corresponds to two overdamped particles confined in parabolic traps with strengths $k_x$ and $k_y$.
The particles interact via a spring with spring constant $\kappa$.
In addition, the parameter $\delta$ describes a nonreciprocal coupling between the two particles. Similar nonreciprocal interactions have recently been realized experimentally in optically levitated particles \cite{rie22}.
Since the dynamics is linear, we can analytically compute the steady state and heat dissipation rates, as well as the power spectral density and response matrix (Appendix B.2).
We explicitly find for subsystem $X$ (sensor),
\begin{subequations}
\begin{align}
\dot{Q}_\text{diss}^X &= \frac{2 T \delta (\delta + \kappa)}{\gamma \mathcal{T}} \label{q-diss}, \\
\frac{T_\text{eff}^X(\omega)}{T} &= 1 + \frac{\dot{Q}_\text{diss}^X}{T} \frac{\gamma \mathcal{T}}{\mathcal{Q}^2 + \kappa^2 - \delta^2 + (\gamma \omega)^2} \label{t-eff}, 
\end{align} \label{linear-sensor}%
\end{subequations}
where we have defined the trace $\text{tr}(\bm{K}) = \mathcal{T}$ and determinant $\det(\bm{K}) = \mathcal{D}$ of the force matrix, as well as $\mathcal{Q} = k_y + \kappa$.
In order to have a stable steady state, we impose the condition $\mathcal{D} > 0$; from the inequality $\mathcal{T} \geq \sqrt{2 \mathcal{D}}$, we also have $\mathcal{T} > 0$.
From \eqref{linear-sensor}, we see that, as expected, a reduced effective temperature is realized when  $\dot{Q}_\text{diss}^X < 0$, that is, for $- \kappa < \delta < 0$.
This implies that both reciprocal ($\kappa \neq 0$) and nonreciprocal ($\delta \neq 0$) couplings are required to decrease the effective temperature.

In order to evaluate the signal-to-noise ratio (\ref{SNR-definition}), we further compute response and  variance:
\begin{align}
\bar{R}_v^X(\omega)^2 &= \frac{\omega^2 \big[\mathcal{Q}^2 + (\gamma \omega)^2 \big]}{\big[(\lambda^+)^2 + (\gamma \omega)^2 \big] \big[(\lambda^-)^2 + (\gamma \omega)^2 \big]} \label{SNR-linear}, \\
\text{Var}_\text{st}(x) &= \frac{T (\gamma \sigma + 2 \mathcal{Q}) - \sqrt{\gamma \sigma \big[ \gamma \sigma  - 4 \frac{ (\mathcal{Q} - \lambda^+) (\mathcal{Q} - \lambda^-)}{\lambda^+ + \lambda^-} \big]} }{2 \lambda^+ \lambda^-} \n ,
\end{align}
where $\lambda^{\pm}$ are the eigenvalues of the force matrix $\bm{K}$.
The response  function does not explicitly depend on the overall dissipation $\sigma = \dot{Q}_\text{diss}/T$, that is, on how far the overall system is driven from equilibrium, contrary to the variance.
 Therefore, for a given response function, the fluctuations can generally be reduced by driving the system out of equilibrium.
Reducing the variance requires $\kappa \neq 0$, that is, just as for the reduction of the effective temperature, both reciprocal and nonreciprocal coupling are necessary to achieve enhanced sensing.

We next numerically illustrate the beneficial role of the demon on the sensor for a small periodic perturbation, $f(t)= \epsilon \cos(\omega_0 t)$, applied to the sensor. To that end, we set the amplitude of the nonequilibrium response at frequency $\omega_0$ to the corresponding equilibrium response, $\bar{R}_v^X(\omega_0) = \bar{R}_{v,\text{eq}}^X(\omega_0)$, where 
$
\bar{R}_{v,\text{eq}}^X(\omega) = \sqrt{{(\gamma \omega)^2}/{k_x^2 + (\gamma \omega)^2}} \label{equilibrium-response}
$
is the response spectrum of the sensor in the absence of the demon.
We also fix the total rate of dissipation $\sigma$.
Then, we numerically minimize the variance with respect to the eigenvalues $\lambda^+$ and $\lambda^-$, which gives us the least possible amount of fluctuations for a given response and dissipation.

Figure 2a) displays the response of the sensor $x(t)$ in equilibrium (gray) and in the presence of the demon (blue) to the small perturbation $f(t)$ (green) for the optimized parameters. A strong decrease in fluctuations is clearly visible; for the considered example, it amounts to a factor of $3.1$ improvement in the signal-to-noise ratio. The behavior of the demon  (purple) is shown in the inset for comparison; as discussed in more details below, it corresponds to an almost unstable mode that exhibits much larger fluctuations than the sensor. Figure 2b) moreover shows the response function $\bar{R}_v^X(\omega)$, \eqref{response-absolute} (black), the signal-to-noise ratio $\text{SNR}^X(\omega)$, \eqref{SNR-definition} (blue),  and the effective temperature $T_\text{eff}^X(\omega)$ (red), relative to their respective equilibrium values, as a function of the frequency $\omega$.
At frequencies lower than the reference frequency $\omega_0$, the response in the presence of the demon is enhanced, both in terms of its absolute value and its real part.
At intermediate frequencies, the coupling to the demon reduces the response, while at high frequencies, where we essentially measure the viscosity of the environment, the response is unaffected by the demon.
Comparing the signal-to-noise ratio to the effective temperature, we see that, while $T_\text{eff}^X(\omega)$ is reduced below the environmental temperature at all frequencies, this reduction may not reflect the observed response of the sensor.

\textit{Fundamental sensing limit.}
To investigate the fundamental limit on the performance of the nonequilibrium sensor, we now consider the results of the optimization as a function of the reference frequency $\omega_0$, as shown in Fig.~\ref{fig-optimization}a).
For frequencies above the characteristic relaxation rate $\omega_\text{c} = k_x/\gamma$ of the sensor, where  response and fluctuations are governed by the properties of the environment rather than the system, no improvement is possible.
By contrast, at low frequencies $\omega_0 \ll \omega_\text{c}$, the signal-to-noise ratio can be significantly enhanced above its equilibrium value.
In particular, in the low frequency limit, where the equilibrium signal-to-noise saturates at a value of unity for the present parameters, the optimized nonequilibrium signal-to-noise ratio diverges as $\omega_0^{-1/4}$.
Likewise, the optimized effective temperature vanishes as $\omega_0^{1/2}$ at low frequencies (Fig.~\ref{fig-optimization}b).
This implies that, for sensing of low-frequency signals, in particular of constant forces, the amount of fluctuations can be decreased arbitrarily, while keeping the response and dissipation finite.
Specifically, using the scaling of the parameters obtained from the numerical minimization, we find for given $\sigma$ and in the limit $\omega_0 \rightarrow 0$ (Appendix B.3),
\begin{align}
\frac{\text{Var}_\text{opt}(x)}{\text{Var}_\text{eq}(x)} \simeq \sqrt{\frac{8 \omega_0}{\sigma}}, \qquad \frac{T_\text{eff}^X(\omega_0)}{T} \simeq \sqrt{\frac{\omega_0}{\sigma}} \label{asymptotic} ,
\end{align}
which agrees with the results obtained by explicit numerical optimization in the low-frequency regime.

To understand the origin of this dramatic improvement, it is useful to consider the optimal values of the eigenvalues $\lambda^{\pm}$.
The increase in the signal-to-noise ratio  is accompanied by a decrease of the smaller eigenvalue $\lambda^- \simeq \gamma \omega_0$.
As a result, the stability of one of the eigenmodes of the system decreases.
Since the response of the sensor is kept fixed, this  eigenmode corresponds to the degree of freedom of the demon, which would be unstable, with negative spring constant, without the stabilizing coupling to the sensor.
This instability allows the demon to \enquote{absorb} the fluctuations of the sensor, thus improving the corresponding signal-to-noise  at the expense of its own  fluctuations, which grow as $\text{Var}_\text{st}(y) \simeq T/(\gamma \omega_0)$ in the low-frequency limit (Fig.~3a, inset).

\textit{Conclusions}. We have investigated the physical limits on nonequilibrium sensing by analyzing a general sensor coupled to a demon. We have shown that in the presence of nonreciprocal coupling, the nonequilibrium action of the demon can significantly suppress fluctuations  while keeping the response unaffected. As a result, the signal-to-noise ratio can be strongly enhanced compared to its equilibrium value. Remarkably, it may even diverge for low frequencies in linear systems, revealing that there is no fundamental limit on nonequilibrium sensing. 
This suggests that appropriately designed nonequilibrium systems might be used for highly accurate sensing even in the presence of large environmental fluctuations.

\textit{Acknowledgements.} AD is supported by JSPS KAKENHI (Grant No. 22K13974 and 24H00833). EL would like to thank the Physics Department at the University of Kyoto, where part of this work was carried out, for their hospitality. He further acknowledges support from the DFG (Grant FOR 2724).

\appendix

\onecolumngrid

\section{Harada-Sasa relation for subsystems}

\subsection{Setup and second law}
We consider the overdamped Langevin equation for a system $Z$ with configuration $\bm{z} \in \mathbb{R}^d$ \cite{ris89}
\begin{align}
\bm{\gamma} \dot{\bm{z}}(t) = \bm{f}(\bm{z}(t)) + \sqrt{2 \bm{\gamma} \kb T} \bm{\xi}(t) \label{langevin-app} .
\end{align}
Here, $\bm{f}(\bm{z})$ describes the forces (conservative and non-conservative external forces, as well as interaction forces) acting in the system, $T$ is the temperature of the environment and $\bm{\gamma}$ is a positive definite $\mathbb{R}^d \times \mathbb{R}^d$ friction matrix.
$\bm{\xi}(t) \in \mathbb{R}^d$ is a vector of mutually independent Gaussian white noises.
Here and in the following we set the Boltzmann constant $\kb = 1$ to simplify the notation.
The steady-state probability density $p(\bm{z})$ of this system is determined by the Fokker-Planck equation
\begin{align}
\grad_z^\text{T} \big( \bm{\nu}(\bm{z}) p(\bm{z}) \big) = 0 \qquad \text{with} \qquad \bm{\nu}(\bm{z}) = \bm{\gamma}^{-1} \big( \bm{f}(\bm{z}) - T \grad_z \ln p(\bm{z}) \big) \label{fpe} .
\end{align}
The local mean velocity $\bm{\nu}^Z(\bm{z})$ also determines the rate of entropy production in the steady state \cite{sei12},
\begin{align}
\sigma = \frac{1}{T} \Av{\bm{\nu}^\text{T} \bm{\gamma} \bm{\nu}} \label{entropy},
\end{align}
where $\Av{\ldots}$ denotes an average with respect to the steady-state probability density.
Note that the entropy production rate vanishes only when the local mean velocity vanishes, which implies the condition
\begin{align}
\bm{f}(\bm{z}) = T \grad_z \ln p(\bm{z}) \equiv -\grad_z U(\bm{z}) \qquad \Rightarrow \qquad p(\bm{z}) = \frac{e^{-\frac{U(\bm{z})}{T}} }{\int d\bm{z} \ e^{-\frac{U(\bm{z})}{T}}} .
\end{align}
That is, the entropy production rate vanishes only for conservative forces, in which case the steady state is an equilibrium state with the corresponding Boltzmann-Gibbs density.
We can derive an equivalent expression for the entropy production rate,
\begin{align}
\sigma = \frac{1}{T} \Av{ \big( \bm{f} - T \grad_z \ln p \big)^\text{T} \bm{\nu}} = \frac{1}{T} \Av{\bm{f}^\text{T} \bm{\nu}} .
\end{align}
Here, we replaced one occurrence of $\bm{\nu}(\bm{z})$ with \eqref{fpe}, and used that $\Av{\grad \psi \bm{\nu}} = 0$ for any gradient field $\bm{\psi}(\bm{z})$ due to \eqref{fpe}.
We can further rewrite this in terms of the Stratonovich product $\circ$ as \cite{sei12}
\begin{align}
\Av{\bm{f}^\text{T} \bm{\nu}} = \Av{\bm{f}^\text{T}(\bm{z}(t)) \circ \dot{\bm{z}}(t)} = \dot{Q}_\text{diss} .
\end{align}
where we identify the rate of heat $\dot{Q}_\text{diss}$ dissipated into the environment with the rate at which work is being done on the system.
Since the entropy production rate is by definition positive, we therefore find the second law of thermodynamics in the steady state
\begin{align}
\dot{Q}_\text{diss} = T \sigma \geq 0 \label{second-law} .
\end{align}
In other words, the system is constantly dissipating heat into the environment, except when it is in equilibrium.

\subsection{Subsystems and information thermodynamics} \label{sec-subsys}
We now divide the degrees of freedom into two subsets, $Z = X + Y$ with $\bm{z} = (\bm{x},\bm{y})$, which we identify with two subsystems $X$ and $Y$, for example, the sensor and the demon.
We further assume that the friction matrix $\bm{\gamma} = \bm{\gamma}^Z$ is block-diagonal
\begin{align}
\bm{\gamma}^Z = \begin{pmatrix} \bm{\gamma}^X & 0 \\ 0 & \bm{\gamma}^Y \end{pmatrix} .
\end{align}
Physically, this condition means that the noise acting on the two subsystems is uncorrelated; it is also referred to as bipartite condition \cite{hor14}.
Under this assumption, the entropy production rate decomposes into two positive contributions
\begin{align}
\sigma = \sigma^X + \sigma^Y \qquad \text{with} \qquad \sigma^X = \frac{1}{T} \Av{\bm{\nu}^{X,\text{T}} \bm{\gamma}^X \bm{\nu}^X} ,
\end{align}
and similar for $\sigma^Y$, where $\bm{\nu}^X(\bm{z})$ denotes the components of the local mean velocity vector corresponding to subsystem $X$.
Repeating the same calculation as for the overall system, we have
\begin{align}
\sigma^X = \frac{1}{T} \Av{ \big( \bm{f}^X - T \grad_x \ln p^Z \big)^\text{T} \bm{\nu}^X}_\text{st},
\end{align}
where $\bm{f}^X(\bm{z})$ are the forces acting on subsystem $X$ and we write $p(\bm{z}) = p^Z(\bm{z}) = p^{X+Y}(\bm{x},\bm{y})$ to clarify that we mean the probability density of the compound system $Z$.
However, since $\grad_x \ln p^Z(\bm{z})$ is generally not a gradient field with respect to $\bm{z}$ (there exists no function $\psi(\bm{x},\bm{y})$ such that $(\grad_x,\grad_y) \psi(\bm{x},\bm{y}) = (\grad_x \ln p^Z(\bm{x},\bm{y}), 0)$) this term does vanish and we have
\begin{align}
\sigma^X = \frac{1}{T} \Av{\bm{f}^{X,\text{T}}(\bm{z}(t)) \circ \dot{\bm{x}}(t)} - l^X = \frac{\dot{Q}_\text{diss}^X}{T} - l^X .
\end{align}
Here, we $\dot{Q}_\text{diss}^X$ is the rate at subsystem $X$ dissipates heat into the environment and the quantity $l^X$ is the so-called learning rate \cite{hor14}
\begin{align}
l^X = \Av{\bm{\nu}^{X,\text{T}} \grad_x \ln p^Z }_\text{st} \label{learning-rate-def}.
\end{align}
Instead of the second law \eqref{second-law}, we now have
\begin{align}
\dot{Q}_\text{diss}^X - T l^X = T \sigma^X \geq 0 \label{second-law-info} .
\end{align}
In particular, if the learning rate $l^X$ is sufficiently negative, we can have $\dot{Q}_\text{diss}^X < 0$, that is, subsystem $X$ continuously absorbs heat from the environment and converts it into work, in apparent violation of the second law of thermodynamics.
However, since we have a similar relation for subsystem $Y$,
\begin{align}
\dot{Q}_\text{diss}^Y - T l^y = T \sigma^Y \geq 0 ,
\end{align}
and the relation $l^X = - l^Y$, the second law \eqref{second-law} is restored for the overall system.
In other words, the apparent negative dissipation of subsystem $X$ is always (over-)compensated by a positive dissipation of the other subsystem $Y$, ensuring that the overall dissipation remains positive.
Let us briefly motivate the term learning rate.
The mutual information between the subsystems $X$ and $Y$ is defined as
\begin{align}
I^{X:Y} = \int d\bm{x} \int d\bm{y} \ \ln \bigg( \frac{p^{X+Y}(\bm{x},\bm{y})}{p^X(\bm{x}) p^Y(\bm{y})} \bigg) p(\bm{x},\bm{y}) = D_\text{KL}(p^{X+Y} \Vert p^X p^Y),
\end{align}
where $p^X(\bm{x}) = \int d\bm{y} \ p^{X+Y}(\bm{x},\bm{y})$ is the marginal probability density of subsystem $X$ and $D_\text{KL}(p,q)$ denotes the Kullback-Leibler divergence between two probability densities $p$ and $q$.
The mutual information is a positive measure of the correlations between $X$ and $Y$; it vanishes only when the two subsystems are independent, $p^{X+Y}(\bm{x},\bm{y}) = p^{X}(\bm{x}) p^Y(\bm{y})$.
For a time-dependent system, we can decompose the change in mutual information into two contributions that can be attributed to the two subsystems
\begin{align}
d_t I^{X:Y} = l^X + l^Y \qquad \text{with} \qquad l^X = \Av{\bm{\nu}^{X,\text{T}} \grad_x \ln p^{X+Y}} + d_t S^X,
\end{align}
where $S^X = - \int d\bm{x} \ \ln(p^X(\bm{x})) p^X(\bm{x})$ is the differential entropy of system $X$.
Therefore, the learning rate $l^X$ quantifies the rate at which the correlations between $X$ and $Y$ increase due to subsystem $X$; it is the rate at which subsystem $X$ is acquiring information (learning) about subsystem $Y$.
In the steady state, the time-derivative of the mutual information as well as of the differential entropy vanish, however, the individual learning rates remain non-zero.
Using that $l^X = -l^Y$ in this case, we can also write \eqref{second-law-info} as
\begin{align}
\dot{Q}_\text{diss}^X + T l^Y \geq 0 \qquad \Rightarrow \qquad \dot{Q}_\text{diss}^X \geq - T l^Y \label{second-law-info-2}.
\end{align}
Thus, a negative rate of dissipation of subsystem $X$ necessarily requires a positive learning rate of subsystem $Y$.
In other words, by acquiring information about $X$, subsystem $Y$ (the \enquote{demon}) can allow $X$ to apparently violate the second law of thermodynamics; at the cost of increasing its own dissipation.

\subsection{Power spectral density}
So far, we only considered the average behavior of the system described by \eqref{langevin-app}.
However, due to its inherent stochasticity, any observable $a(\bm{z}(t))$ measured in the system $Z$ is subject to the fluctuations of $\bm{z}(t)$.
One way of characterizing these fluctuations is in terms of the power-spectral density (PSD).
This can be defined by considering the finite-time Fourier transform of the observable \cite{ris89},
\begin{align}
\hat{a}_\tau(\omega) = \int_0^\tau dt \ e^{i \omega t} a(\bm{z}(t)) .
\end{align}
The PSD is then defined as the long-time limit of the fluctuations of $\hat{a}_\tau(\omega)$,
\begin{align}
S_a(\omega) = \lim_{\tau \rightarrow \infty} \frac{1}{\tau} \Big( \Av{\vert \hat{a}_\tau(\omega) \vert^2} - \vert \Av{\hat{a}_\tau(\omega)} \vert^2 \Big) .
\end{align}
The Wiener-Khinchin theorem states that this is equal to the Fourier-transform of the steady-state autocorrelation of $a(\bm{z}(t))$,
\begin{align}
S_a(\omega) = \int_{-\infty}^{\infty} dt \ e^{i \omega t} \ \Big( \Av{a(\bm{z}(t)) a(\bm{z}(0))} - \Av{a}^2 \Big) = 2 \int_0^\infty dt \ \cos(\omega t) \Big( \Av{a(\bm{z}(t)) a(\bm{z}(0))} - \Av{a}^2 \Big) .
\end{align}
Note that, since the autocorrelation is symmetric in time, the PSD is real and symmetric in $\omega$; in fact, it is positive as is clear from its definition.
Using Parseval's identity, we further have for the integral over all frequencies,
\begin{align}
\frac{1}{\pi} \int_0^\infty d\omega \ S_a(\omega) = \Av{a^2} - \Av{a}^2 = \lim_{t \rightarrow 0} \Big( \Av{a(\bm{z}(t)) a(\bm{z}(0))} - \Av{a}^2 \Big) \label{psd-short-time} ,
\end{align}
that is, the steady-state fluctuations of $a(\bm{z})$.
Generally, the frequency integral over the PSD is determined by the short-time behavior of the autocorrelation function.
We can similarly define the cross-PSD for two different observables $a(\bm{z}(t))$ and $b(\bm{z}(t))$,
\begin{align}
S_{a,b}(\omega) = \int_{0}^{\infty} dt \ \cos(\omega t) \Big(  \Av{a(\bm{z}(t)) b(\bm{z}(0))} + \Av{b(\bm{z}(t)) a(\bm{z}(0)} - 2 \Av{a} \Av{b} \Big) .
\end{align}
With the identification $S_{a}(\omega) = S_{a,a}(\omega)$, the PSD matrix
\begin{align}
\bm{S}_{a,b}(\omega) = \begin{pmatrix} S_{a,a}(\omega) & S_{a,b}(\omega) \\ S_{a,b}(\omega) & S_{b,b}(\omega) \end{pmatrix}
\end{align}
is a positive definite matrix, whose frequency integral is equal to the steady-state covariance matrix of $a(\bm{z})$ and $b(\bm{z})$.
This can be extended to an arbitrary number of observables, in particular, choosing the entries of $\bm{z}$ itself as observables, we can define the PSD matrix of $\bm{z}$,
\begin{align}
\big(\bm{S}^Z(\omega) \big)_{kl} = S_{z_k,z_l}(\omega) .
\end{align}
Recalling the decomposition into subsystems $Z = X+Y$, we can also write this as
\begin{align}
\bm{S}^Z(\omega) = \begin{pmatrix} \bm{S}^X(\omega) & \bm{S}^{XY}(\omega) \\ \bm{S}^{XY,\text{T}}(\omega) & \bm{S}^Y(\omega) \end{pmatrix} \label{PSD-subsystem},
\end{align}
where $\bm{S}^X(\omega)$ is the PSD matrix of subsystem $X$.
We remark that we can also define the corresponding PSD matrix for the velocity, i.~e.~
\begin{align}
\big(\bm{S}_v^Z(\omega) \big)_{kl} = \int_{0}^\infty dt \ \cos(\omega t) \ \bigg( \Big( \Av{\dot{z}_k(t) \dot{z}_l(0)} + \Av{\dot{z}_l(t) \dot{z}_k(0)} \Big) - 2 \Av{\dot{z}_k} \Av{\dot{z}_l} \bigg) \label{velocity-PSD} .
\end{align}
This is related to the PSD matrix of $Z$ via
\begin{align}
\bm{S}_v^Z(\omega) = \omega^2 \bm{S}^Z(\omega) .
\end{align}

\subsection{Response and Harada-Sasa relation}
We now consider applying a small time-dependent perturbation $\epsilon \bm{\phi}(t)$ to the system described by \eqref{langevin-app},
\begin{align}
\bm{\gamma} \dot{\bm{z}}(t) = \bm{f}(\bm{z}(t)) + \epsilon \bm{\phi}(t) + \sqrt{2 \bm{\gamma} T} \bm{\xi}(t) 
\end{align}
where $\epsilon \ll 1$ is a small parameter.
We assume that we have $\bm{\phi}(t) = 0$ for $t < 0$ and that the system is initially in the steady state.
We want to characterize the response of the system to this perturbation, which we describe in terms of the difference
\begin{align}
\Av{\bm{z}(t)}^\epsilon - \Av{\bm{z}},
\end{align}
that is, how much the expected value of $\bm{z}$ at time $t$ changes relative to its steady-state value as a result of the perturbation.
In the linear response regime, $\epsilon \rightarrow 0$, we expect this change to be proportional to $\epsilon$ and write it as
\begin{align}
\Av{\bm{z}(t)}^\epsilon - \Av{\bm{z}} = \int_0^t dt' \ \Av{\dot{\bm{z}}(t')}^\epsilon \simeq \epsilon \int_0^t dt' \int_0^{t'} dt'' \ \bm{\mathcal{R}}^Z_v(t'-t'') \bm{\phi}(t'') + O(\epsilon^2) \label{response-definition} .
\end{align}
The matrix $\bm{\mathcal{R}}^Z_v(t'-t'')$ is the velocity response matrix, which describes the change in the velocity at time $t'$ as a consequence of the applied perturbation force at time $t'' < t'$; specifically, $\mathcal{R}^Z_{v,kl}$ is the change of the velocity in direction $k$ to an applied force in direction $l$.
In general, we can express the response of an observable to the perturbation as a correlation function between the observable and some other quantity in the unperturbed system (see Refs.~\citep{spe06,har06}),
\begin{align}
\Av{\bm{z}(t)}^\epsilon - \Av{\bm{z}}^0 \simeq \epsilon \Av{\bm{z}(t) Q(t)}^0,
\end{align}
where $\av{\ldots}^0$ means the expectation with respect to the unperturbed dynamics.
The quantity $Q(t)$ is the derivative of the path-probability density with respect to $\epsilon$,
\begin{align}
Q(t) = \partial_\epsilon \ln \mathbb{P}^\epsilon(\hat{\bm{z}}) \Big\vert_{\epsilon = 0} = \frac{1}{2 T} \int_0^t dt'' \ \bm{\phi}^\text{T}(t'') \big( \dot{\bm{z}}(t'') - \bm{\gamma}^{-1} \bm{f}(\bm{z}(t'') \big) = \frac{1}{2 T}  \int_0^t dt'' \   \bm{\phi}^\text{T}(t'') \sqrt{2 T \bm{\gamma}^{-1}} \bm{\xi}(t'') ,
\end{align}
where we used \eqref{langevin-app}.
We find
\begin{align}
\Av{\bm{z}(t)}^\epsilon - \Av{\bm{z}} = \frac{1}{2 T} \int_0^t dt' \int_0^t dt'' \ \Av{\dot{\bm{z}}(t') \Big( \bm{\phi}^\text{T}(t'') \sqrt{2 T \bm{\gamma}^{-1}} \bm{\xi}(t'') \Big) }
\end{align}
Since the noise at later times $t'' > t'$ is independent of the velocity at time $t'$, this part of the integral does not contribute, and we have
\begin{align}
\Av{\bm{z}(t)}^\epsilon - \Av{\bm{z}} = \frac{1}{2 T} \int_0^t dt' \int_0^{t'} dt'' \ \Av{\dot{\bm{z}}(t') \Big( \bm{\phi}^\text{T}(t'') \big( \dot{\bm{z}}(t'') - \bm{\gamma}^{-1} \bm{f}(\bm{z}(t'') \big) \Big) },
\end{align}
which reflects causality, i.~e.~the perturbation only affects the velocity at a later time.
Comparing this to \eqref{response-definition}, we can identify the response matrix
\begin{align}
\bm{\mathcal{R}}_v^Z(t'-t'') = \frac{1}{2 T} \Av{\dot{\bm{z}}(t') \big( \dot{\bm{z}}(t'') - \bm{\gamma}^{-1} \bm{f}(\bm{z}(t'')) \big)^\text{T}}
\end{align}
The first term is precisely the velocity correlation function, that is, we have for $t' \geq t''$,
\begin{align}
\bm{\mathcal{R}}_v^Z(t'-t'') = \frac{1}{2 T} \Big( \Av{\dot{\bm{z}}(t') \dot{\bm{z}}^\text{T}(t'')} - \Av{\dot{\bm{z}}(t') \bm{f}^\text{T}(\bm{z}(t''))} \bm{\gamma}^{-1} \Big) .
\end{align}
Let us now consider the following expression
\begin{align}
\text{tr}\bigg( \bm{\gamma} \Big( \Av{\dot{\bm{z}}(t) \dot{\bm{z}}^\text{T}(0)} - 2 T \bm{\mathcal{R}}_v^Z(t) \Big) \bigg) \label{response-trace},
\end{align}
where tr denotes the trace.
Plugging in the above expression for the response matrix, we obtain
\begin{align}
\text{tr}\bigg( \bm{\gamma} \Big( \Av{\dot{\bm{z}}(t) \dot{\bm{z}}^\text{T}(0)} - 2 T \bm{\mathcal{R}}_v^Z(t) \Big) \bigg) = \text{tr} \bigg( \bm{\gamma} \Av{\dot{\bm{z}}(t) \bm{f}^\text{T}(\bm{z}(0))} \bm{\gamma}^{-1} \bigg) = \Av{\bm{f}^\text{T}(\bm{z}(0)) \dot{\bm{z}}(t)} ,
\end{align}
where we used the invariance of the trace under cyclic permutations.
Multiplying by $\cos(\omega t)$ and integrating over $t$, we get
\begin{align}
\text{tr}\bigg( \bm{\gamma} \Big( \bm{S}^z_v(\omega) - 2 T \bm{R}_v^Z(\omega) \Big) \bigg) = \int_0^\infty dt \ \cos(\omega t) \Av{\bm{f}^\text{T}(\bm{z}(0)) \dot{\bm{z}}(t)} ,
\end{align}
where we identified the velocity PSD matrix \eqref{velocity-PSD}.
We also have for the the Fourier transform of the response matrix,
\begin{align}
\bm{R}_v^Z(\omega) + i \tilde{\bm{R}}_v^Z(\omega) = \int_{-\infty}^\infty dt \ e^{i \omega t} \bm{\mathcal{R}}_v^Z(t) = \int_0^\infty dt \ e^{i \omega t} \bm{\mathcal{R}}_v^Z(t) = \int_0^\infty dt \ \cos(\omega t) \bm{\mathcal{R}}_v^Z(t) + i \int_0^\infty dt \ \sin(\omega t) \bm{\mathcal{R}}_v^Z(t) ,
\end{align}
since the response function vanishes for negative arguments due to causality.
Finally, using that the integral over all frequencies is is equal to the short-time behavior of the integrand, or, more formally, the relation
\begin{align}
\int_0^\infty d\omega \ \cos(\omega t) = \pi \delta(t),
\end{align}
we obtain the identity
\begin{align}
\frac{1}{\pi} \int_0^\infty d\omega \ \text{tr} \bigg( \bm{\gamma} \Big( \bm{S}^Z_v(\omega) - 2 T \bm{R}_v^Z(\omega) \Big) \bigg) = \Av{\bm{f}^\text{T} \circ \dot{\bm{z}}} = \dot{Q}_\text{diss} \label{harada-sasa} .
\end{align}
This is known as the Harada-Sasa relation \cite{har05,har06}; it expresses the dissipated heat as an integral over the violation of the fluctuation-dissipation theorem.
Since the friction matrix is assumed as block-diagonal, we can decompose the expression in \eqref{response-trace} into contributions due to $X$ and $Y$,
\begin{align}
\text{tr}\bigg( \bm{\gamma} \Big( \Av{\dot{\bm{z}}(t) \dot{\bm{z}}^\text{T}(0)} - 2 T \bm{\mathcal{R}}_v^Z(t) \Big) \bigg) &= \text{tr}\bigg( \bm{\gamma}^X \Big( \Av{\dot{\bm{x}}(t) \dot{\bm{x}}^\text{T}(0)} - 2 T \bm{\mathcal{R}}_v^X(t) \Big) \bigg) + \text{tr}\bigg( \bm{\gamma}^Y \Big( \Av{\dot{\bm{y}}(t) \dot{\bm{y}}^\text{T}(0)} - 2 T \bm{\mathcal{R}}_v^y(t) \Big) \bigg) \nn
& = \Av{\bm{f}^{X,\text{T}}(0) \dot{\bm{x}}(t)} + \Av{\bm{f}^{Y,\text{T}}(0) \dot{\bm{y}}(t)},
\end{align}
where we write the response matrix in block form similar to the PSD matrix in \eqref{PSD-subsystem},
\begin{align}
\bm{\mathcal{R}}^Z(t) = \begin{pmatrix} \bm{\mathcal{R}}^X(t) & \bm{\mathcal{R}}^{XY}(t) \\ \bm{\mathcal{R}}^{YX}(t) & \bm{\mathcal{R}}^Y(t) \end{pmatrix} \label{response-subsystem}.
\end{align}
The matrix $\bm{\mathcal{R}}^X(t)$ describes the response of subsystem $X$ to a perturbation applied to $X$, while $\bm{\mathcal{R}}^{XY}(t)$ describes the response of $X$ to a perturbation applied to $Y$.
Note that, in contrast to the PSD matrix the response matrix is generally not symmetric, $\bm{\mathcal{R}}^{YX}(t) \neq \bm{\mathcal{R}}^{XY,\text{T}}(t)$.
Repeating the same calculation as above, we therefore find for subsystem $X$,
\begin{align}
\frac{1}{\pi} \int_0^\infty d\omega \ \text{tr} \bigg( \bm{\gamma}^X \Big( \bm{S}^X_v(\omega) - 2 T \bm{R}_v^X(\omega) \Big) \bigg) = \Av{\bm{f}^{X,\text{T}} \circ \dot{\bm{x}}} = \dot{Q}^X_\text{diss} \label{harada-sasa-subsystem} .
\end{align}
Thus, the Harada-Sasa relation also holds separately for each subsystem---the violation of the fluctuation-dissipation theorem of subsystem $X$ is equal to the heat dissipated by subsystem $X$.
As discussed before, the latter can be negative under suitable conditions, which implies that, compared to an equilibrium system, the response can be enhanced relative to the fluctuations.

\section{Response and fluctuations for linear systems}

\subsection{General relations}
Let us now consider a particular case of \eqref{langevin-app} with a linear force
\begin{align}
\bm{\gamma} \dot{\bm{z}}(t) = - \bm{K} \bm{z}(t) + \sqrt{2 \bm{\gamma} T} \bm{\xi}(t) .
\end{align}
Here, the force matrix $\bm{K}$ is assumed to have eigenvalues with strictly positive real parts, so that the system has a stable steady state.
In this case, we can determine the steady-state explicitly; it is given by the Gaussian probability density
\begin{align}
p^Z_\text{st}(\bm{z}) = \frac{1}{(2 \pi)^d \det(\bm{\Xi}^Z)} \exp \bigg(- \frac{1}{2} \bm{z}^\text{T} {\bm{\Xi}^Z}^{-1} \bm{z} \bigg),
\end{align}
whose covariance matrix is the solution of the Lyapunov equation
\begin{align}
\bm{\gamma}^{-1} \bm{K} \bm{\Xi}^Z + \bm{\Xi}^Z \bm{K}^\text{T} \bm{\gamma}^{-1} = 2 T \bm{\gamma}^{-1} \label{lyapunov} .
\end{align}
To keep the notation compact, we define the matrices $\bm{A} = \bm{\gamma}^{-1} \bm{K}$ and $\bm{B} = \bm{\gamma}^{-1} T$, so that
\begin{align}
\bm{A} \bm{\Xi}^Z + \bm{\Xi}^Z \bm{A}^\text{T} = 2 \bm{B} .
\end{align}
The local mean velocity is given by
\begin{align}
\bm{\nu}(\bm{z}) = \big( - \bm{A} + \bm{B} {\bm{\Xi}^Z}^{-1} \big) \bm{z},
\end{align}
and the entropy production rate by
\begin{align}
\sigma = \text{tr} \Big( \bm{B}^{-1} \big( - \bm{A} + \bm{B} {\bm{\Xi}^Z}^{-1} \big) \bm{\Xi}^Z \big( - \bm{A} + \bm{B} {\bm{\Xi}^Z}^{-1} \big)^\text{T} \Big) = \frac{1}{T} \text{tr} \Big( \bm{\gamma}^{-1} \big( \bm{K} \bm{\Xi}^Z \bm{K}^\text{T} - \bm{K} \big) \Big),
\end{align}
where we used \eqref{lyapunov} in the second equality.
The system is in equilibrium if the the force matrix $\bm{K}$ is symmetric, in which case we have $\bm{\Xi}^Z = T \bm{K}^{-1}$ and thus $\bm{A} = \bm{B} {\bm{\Xi}^Z}^{-1}$ and $\sigma = 0$.
The average of $\bm{z}$ at time $t$, conditioned on a value $\bm{z}'$ at time $0$ is given by
\begin{align}
\Av{\bm{z}(t) \vert \bm{z}'} = e^{- \bm{A} t} \bm{z}',
\end{align}
which allows us to compute the correlation function
\begin{align}
\Av{\bm{z}(t) \bm{z}^\text{T}(0)} = \Av{ \Av{\bm{z}(t) \vert \bm{z}'} {\bm{z}'}^{\text{T}}} = e^{- \bm{A} t} \bm{\Xi}^Z .
\end{align}
Using this, we can evaluate the velocity PSD matrix
\begin{align}
\bm{S}_v^Z(\omega) = \omega^2 \int_0^\infty dt \ \cos(\omega t) \Big( e^{- \bm{A} t} \bm{\Xi}^Z + \bm{\Xi}^Z e^{-\bm{A}^\text{T} t} \Big) =\omega^2 \Big( \big(\bm{A}^2 + \omega^2 \bm{I} \big)^{-1} \bm{A} \bm{\Xi}^Z + \bm{\Xi}^Z \bm{A}^\text{T} \big( \bm{A}^2 + \omega^2 \bm{I} \big)^{-1,\text{T}} \Big),
\end{align}
and the Fourier-transformed velocity response matrix is given by
\begin{align}
\bm{R}_v^Z(\omega) + i \tilde{\bm{R}}_v^Z(\omega) = \bm{\gamma}^{-1} \big(\omega^2 \bm{I}  + i \omega \bm{A} \big) \big( \bm{A}^2 + \omega^2 \bm{I} \big)^{-1},
\end{align}
where $\bm{I}$ denotes the identity matrix.

As in Sec.~\ref{sec-subsys}, we now decompose the system into two subsystems, $Z = X+Y$ with $\bm{z} = (\bm{x},\bm{y})$, where we interpret $X$ as the sensor and $Y$ as an auxiliary system (\enquote{demon}) that we engineer to enhance the performance of $X$.
From \eqref{learning-rate-def}, we can write the learning rate of subsystem $X$ as
\begin{align}
l^X = \Av{\bm{\nu}^{X,\text{T}} \grad_x \ln p^Z} = - \Av{ \grad_x^\text{T} \bm{\nu}^X} = \text{tr} \Big(\big( \bm{A} - \bm{B} {\bm{\Xi}^Z}^{-1} \big)^{X} \Big) = \text{tr} \bigg( {\bm{\gamma}^X}^{-1} \Big(\bm{K}^{X} - T \big({\bm{\Xi}^Z}^{-1} \big)^{X} \Big) \bigg),
\end{align}
where $\bm{K}^{X}$ denotes the upper left block of the matrix $\bm{K}$, corresponding to subsystem $X$.
Note that this expression involves the upper-left block of the inverse covariance matrix of the compound system, which includes correlations between $X$ and $Y$.
Specifically, using the block-inversion formula for matrices, we have
\begin{align}
\big({\bm{\Xi}^Z}^{-1} \big)^{X} = \big( \bm{\Xi}^X - \bm{\Xi}^{XY} {\bm{\Xi}^Y}^{-1} {\bm{\Xi}^{XY}}^\text{T} \big)^{-1} .
\end{align}
The entropy production rate of $X$ is given by
\begin{align}
\sigma^X = \text{tr} \Bigg( {\bm{\gamma}^X}^{-1} \bigg( \frac{1}{T} \bm{K} \bm{\Xi}^Z \bm{K}^\text{T} - 2 \bm{K} + T {\bm{\Xi}^Z}^{-1} \bigg)^X \Bigg) 
\end{align}
and the dissipation rate by
\begin{align}
\dot{Q}_\text{diss}^X = T \big( \sigma^X + l^X \big) = \text{tr} \Big( {\bm{\gamma}^X}^{-1} \big( \bm{K} \bm{\Xi}^Z \bm{K}^\text{T} - T \bm{K} \big)^{X}  \Big).
\end{align}

\subsection{Solvable two-dimensional model}
While the above relations hold for general linear dynamics, an explicit evaluation requires solving the Lyapunov equation \eqref{lyapunov}.
We therefore focus on the two-dimensional case, where this can be done explicitly.
Specifically, we set
\begin{align}
\bm{K} = \begin{pmatrix} k_x + \kappa & -\kappa - \delta \\ - \kappa + \delta & k_y + \kappa \end{pmatrix}, \qquad \bm{\gamma} = \gamma \bm{I} .
\end{align}
This describes two linearly interacting overdamped degrees of freedom $x$ and $y$.
$k_x$ and $k_y$ are the corresponding one-body force constants, $\kappa$ describes a reciprocal interaction and $\delta$ a non-reciprocal interaction.
The system is out of equilibrium if $\delta \neq 0$.
We define the constants
\begin{align}
\mathcal{T} = \text{tr}(\bm{K}) = k_x + k_y + 2 \kappa, \qquad \mathcal{D} = \det(\bm{K}) = k_x k_y + (k_x + k_y) \kappa + \delta^2, \qquad \mathcal{Q} = k_y + \kappa .
\end{align}
In terms of these, the eigenvalues of the force matrix $\bm{K}$ can be written as
\begin{align}
\lambda^{\pm} = \frac{1}{2} \Big( \mathcal{T} \pm \sqrt{\mathcal{T}^2 - 4 \mathcal{D}} \Big)
\end{align}
The requirement for a stable steady state thus implies $\mathcal{T} > 0$ and $\mathcal{D} > 0$.
The resulting covariance matrix is given by
\begin{align}
\bm{\Xi} = \frac{T}{\mathcal{T} \mathcal{D}} \begin{pmatrix} k_y^2 + 3 k_y \kappa + k_x (k_y + \kappa) + 2 (\delta^2 + \delta \kappa + \kappa^2) & k_x (\kappa + \delta) + k_y (\kappa - \delta) + 2 \kappa^2 \\ k_x (\kappa + \delta) + k_y (\kappa - \delta) + 2 \kappa^2 & k_x^2 + 3 k_y \kappa + k_y(k_x+\kappa) + 2 (\delta^2  - \delta \kappa + \kappa^2) \end{pmatrix} .
\end{align}
Using this, we can write the dissipation rate of subsystem $X$, its velocity-PSD and velocity-response as
\begin{gather}
\dot{Q}_\text{diss}^X = \frac{2 T \delta (\delta + \kappa)}{\gamma \mathcal{T}}, \qquad S_v^X(\omega) = \frac{2 T}{\gamma} \frac{(\gamma \omega)^2 \big( \mathcal{Q}^2 + (\delta + \kappa)^2 + (\gamma \omega)^2 \big)}{\mathcal{D}^2 + \big( \mathcal{T}^2 - 2 \mathcal{D} \big) (\gamma \omega)^2 + (\gamma \omega)^4} \\
R_v^X(\omega) = \frac{1}{\gamma} \frac{(\gamma \omega)^2 \big( \mathcal{Q}^2 + \kappa^2 - \delta^2 + (\gamma \omega)^2 \big)}{\mathcal{D}^2 + \big( \mathcal{T}^2 - 2 \mathcal{D} \big) (\gamma \omega)^2 + (\gamma \omega)^4} \qquad \tilde{R}_v^X(\omega) = \frac{1}{\gamma} \frac{(\gamma \omega) \big(\mathcal{Q} \mathcal{D} + (k_x + \kappa) (\gamma \omega)^2 \big)}{\mathcal{D}^2 + \big( \mathcal{T}^2 - 2 \mathcal{D} \big) (\gamma \omega)^2 + (\gamma \omega)^4} \n .
\end{gather}
In equilibrium, that is, for $\delta = 0$, we verify the fluctuation-dissipation theorem,
\begin{align}
S_{v,\text{eq}}^{X}(\omega) = 2 T R_{v,\text{eq}}^{X}(\omega) .
\end{align}
Out of equilibrium, the ratio of the PSD and real part of the response can be used to define the effective temperature
\begin{align}
\frac{T_\text{eff}^X(\omega)}{T} = \frac{S_v^X(\omega)}{2 T R_v^X(\omega)} = \frac{ \mathcal{Q}^2 + (\delta + \kappa)^2 + (\gamma \omega)^2}{ \mathcal{Q}^2 + \kappa^2 - \delta^2 + (\gamma \omega)^2} = 1 + \frac{2 \delta (\delta + \kappa)}{\mathcal{Q}^2 + \kappa^2 - \delta^2 + (\gamma \omega)^2} = 1 + \frac{\gamma \mathcal{T} \dot{Q}_\text{diss}^X }{T \big(\mathcal{Q}^2 + \kappa^2 - \delta^2 + (\gamma \omega)^2 \big)} .
\end{align}
This relation implies that, as expected from \eqref{harada-sasa-subsystem}, an effective temperature that is lower than the environmental temperature (and thus enhanced response relative to the fluctuations) is only possible when $\dot{Q}_\text{diss}^X$ is negative.
From \eqref{second-law-info-2}, this also implies that the subsystem $Y$ has to continuously acquire information about $X$ in order to facilitate the reduction in effective temperature.

\subsection{Optimization of the signal-to-noise ratio}
As in the main text, we introduce the signal-to-noise ratio (SNR) of system $X$,
\begin{align}
\text{SNR}^X(\omega) = \frac{\bar{R}_v^X(\omega) f}{\omega \sqrt{\Xi^X}} .
\end{align}
Here, $\bar{R}_v^X(\omega) = \sqrt{R_v^X(\omega)^2 + \tilde{R}_v^X(\omega)^2}$ is the magnitude of the velocity response, which is related to the response of the coordinate $x$ as $\bar{R}_v^X(\omega) = \omega \bar{R}^X(\omega)$.
$f$ is the magnitude of the perturbation force applied to $X$ and $\Xi^X = \text{Var}_\text{st}(x)$ denotes the steady-state fluctuations of $x$.
Our goal is to maximize the SNR maintaining the magnitude of the response.
We rewrite the response function and fluctuations using the eigenvalues $\lambda^{\pm}$ of the force matrix, as well as the entropy production rate $\sigma$,
\begin{align}
\bar{R}_v^X(\omega)^2 &= \frac{\omega^2 \big(\mathcal{Q}^2 + (\gamma \omega)^2 \big)}{\big((\lambda^+)^2 + (\gamma \omega)^2 \big) \big((\lambda^-)^2 + (\gamma \omega)^2 \big)}, \qquad 
\Xi^X = \frac{T (\gamma \sigma + 2 \mathcal{Q}) - \sqrt{\gamma \sigma \big( \gamma \sigma  - 4 \frac{ (\mathcal{Q} - \lambda^+) (\mathcal{Q} - \lambda^-)}{\lambda^+ + \lambda^-} \big)} }{2 \lambda^+ \lambda^-} \label{response-eigenvalue} .
\end{align}
We see that fixing the response function for all frequencies also determines the parameters $\lambda^{\pm}$ and $\mathcal{Q}$, while the entropy production rate $\sigma$ only enters the fluctuations.
Maximizing the SNR at a given response therefore corresponds to minimizing $\Xi^X$ with respect to $\sigma$.
The minimal value is attained in the limit $\sigma \rightarrow \infty$, that is, when driving the system far from equilibrium.
We find
\begin{align}
\frac{\Xi^X_\text{min}}{\Xi^X_\text{eq}} = \frac{k_{x,\text{eq}} + \frac{k_{y,\text{eq}} \kappa_{\text{eq}}}{k_{y,\text{eq}} + \kappa_{\text{eq}}} + k_{y,\text{eq}} + \kappa_{\text{eq}}}{k_{x,\text{eq}} + k_{y,\text{eq}} + 2\kappa_{\text{eq}} } < 1 .
\end{align}
That is, for a given equilibrium system with parameters $k_{x,\text{eq}}$, $k_{y,\text{eq}}$ and $\kappa_{\text{eq}}$, we can reduce the fluctuations of $x$ and thus improve the SNR by introducing a non-reciprocal coupling and driving the system out of equilibrium, while keeping the response of the system at all frequencies unaffected.
Reducing the variance requires $\kappa \neq 0$, that is, just as for the reduction of the effective temperature, both reciprocal and non-reciprocal coupling are necessary to achieve enhanced sensing.

In practice, however, specifying the entire response function is often too restrictive, since we are rather interested in the response of the sensor at a specific frequency $\omega_0$.
We therefore specify the amplitude of the response $\bar{R}^X(\omega_0) = \bar{R}_{\text{eq}}^X(\omega_0) \equiv R_0$, where 
\begin{align}
\bar{R}_{\text{eq}}^X(\omega) = \sqrt{\frac{1}{k_{x,\text{eq}}^2 + (\gamma \omega)^2}} \label{equilibrium-response-app}
\end{align}
is the response spectrum of the sensor in the absence of the demon.
We remark that, in equilibrium, we have $\Xi^X_\text{eq} = T/k_{x,\text{eq}}$.
Thus, a decrease in the fluctuations of $X$ can only be achieved by increasing the force constant $k_{x,\text{eq}}$, which, however, also decreases the response \eqref{equilibrium-response-app}.
Consequently, if we want to reduce the fluctuations while maintaining the response, we need to drive the system out of equilibrium.
In the following, we therefore also specify the total rate of dissipation $\sigma$, which characterizes how far the overall system is out of equilibrium.
Then, we minimize the variance with respect to the eigenvalues $\lambda^+$ and $\lambda^-$, which gives us the least possible amount of fluctuations for a given response and dissipation.
Since the corresponding optimization problem is non-linear with equality (on the response and dissipation) and inequality (on the eigenvalues, $\lambda^+,\lambda^- >0$) constraints, we carry out the minimization numerically using the \texttt{NMinimize} command of Mathematica.
Since at high frequencies, the response and fluctuations of the system are determined by the environment and we therefore cannot expect a significant enhancement of the response, we focus on the low-frequency limit $\omega_0 \rightarrow 0$.
Numerically, we observe the scaling $\lambda^+ \propto 1/\sqrt{\omega_0}$ and $\lambda^- \propto \omega_0$ for the optimal eigenvalues in the limit $\omega_0 \rightarrow 0$.
We therefore set
\begin{align}
\lambda^+ = \frac{c_1}{\sqrt{\omega_0}} \qquad \text{and} \qquad \lambda^- = c_2 \omega_0,
\end{align}
where $c_1$ and $c_2$ are positive constants.
Plugging this into \eqref{response-eigenvalue} and expanding for small $\omega_0$, we obtain
\begin{align}
\Xi^X \simeq \bigg(\frac{T}{c_1} + \frac{c_1 (\gamma^2 + c_2^2) R_0^2 T}{c_2 \gamma \sigma} \bigg) \sqrt{\omega_0} + O(\omega_0) .
\end{align}
Minimizing the coefficient with respect to $c_1$ and $c_2$, we find the minimal fluctuations
\begin{align}
\Xi^X \simeq 2 \sqrt{2} T R_0 \sqrt{\frac{\omega_0}{\sigma}} + O(\omega_0) .
\end{align}
The covariance matrix is given by
\begin{align}
\bm{\Xi} \simeq T \begin{pmatrix} 2 R_0 \sqrt{\frac{2 \omega_0}{\sigma}} & \Big(\frac{2 R_0^2}{\gamma^2 \sigma \omega_0} \Big)^{\frac{1}{4}} \\ \Big(\frac{2 R_0^2}{\gamma^2 \sigma \omega_0} \Big)^{\frac{1}{4}} & \frac{1}{\gamma \omega_0} \end{pmatrix} .
\end{align}
We see that, for a low-frequency perturbation, the fluctuations of the sensor can be made arbitrarily small, vanishing as $\omega_0^{1/2}$, while maintaining a finite dissipation rate and a finite response to the perturbation.
At the same time, the correlations between the sensor and the demon diverge as $\omega_0^{-1/4}$, while the fluctuations of the demon diverge even faster, as $\omega_0^{-1}$.
This is corroborated by the eigenvectors of the force matrix,
\begin{align}
\bm{e}^- \simeq \begin{pmatrix} \Big( \frac{2 R_0^2 \gamma^2 \omega_0^3}{\sigma} \Big)^{\frac{1}{4}} \\ 1 \end{pmatrix}, \qquad \bm{e}^+ \simeq \begin{pmatrix} - \Big( \frac{1}{2 R_0^2 \gamma^2 \sigma \omega_0} \Big)^{\frac{1}{4}} \\ 1 \end{pmatrix} .
\end{align}
As the smaller eigenvalue $\lambda^-$ vanishes, the corresponding eigenvector $\bm{e}^-$ is oriented in the direction of the demon; the dynamics of the demon become asymptotically unstable, leading to large fluctuations.
By contrast, the larger eigenvalue $\lambda^+$ increases; its eigenvector is oriented in the direction of the sensor.
The increase in $\lambda^+$ therefore stabilizes the sensor, decreasing its fluctuations.

Finally, we can also consider the learning rate and rate of heat dissipation, which scale as
\begin{align}
Q_\text{diss}^X \simeq - l^Y \simeq - \sqrt{\sigma \omega_0} + O(\omega_0)
\end{align}
in the low-frequency limit.
The entropy production rate associated with the sensor $X$ scales as
\begin{align}
\sigma^X \simeq 2 \sqrt{2} \gamma R_0 \sqrt{\sigma} \omega_0^{\frac{3}{2}} + O(\omega_0^2) .
\end{align}
Thus, the information acquired by the demon per period of the driving, $l^Y/\omega_0$, diverges as $\omega_0^{-1/2}$ in the low-frequency limit; in order to suppress the fluctuations of the sensor, the demon has to learn about the fluctuations of the sensor and subsequently dissipate the acquired information in the form of heat into the environment.
On the other hand, the dissipation from the sensor per period of the driving $\sigma^X/\omega_0$ vanishes as $\omega_0^{1/2}$---in the low-frequency limit, the entire dissipation is due to the dynamics of the demon.

\end{document}